\documentclass[conference,letter]{IEEEtran}
\IEEEoverridecommandlockouts

\usepackage{url}
\usepackage{xcolor}
\usepackage{subfigure}
\usepackage{alltt}
\usepackage{fancyhdr}
\usepackage{amstext}
\usepackage[cmex10]{amsmath} 
\usepackage{amssymb}  
\usepackage{amsthm}
\usepackage{amsfonts}
\usepackage{mathtools}
\usepackage{stmaryrd} 
\usepackage{newfloat} 
\usepackage{framed}   
\usepackage{cite}
\usepackage{graphicx}
\usepackage{textcomp}
\usepackage{xcolor}
\usepackage{dsfont} 
\usepackage{upgreek}   
\usepackage{enumitem}  
\usepackage{marvosym}  

\makeatletter
\newcommand{\stackover}{\genfrac{}{}{0pt}{}}
\makeatother
\def\padlock{\stackover{%
               \raisebox{-3pt}[0.8ex][0pt]{$\Cap$}%
            }{%
               \raisebox{0pt}[1.2ex][0pt]{$\square$}%
            }%
            }

\def\flat{{\rm E}\kern-2.7pt\smash{\rm P}}
\def\AT{\forall\kern-1pt\smash{\rm I}}

\makeatletter
\@ifpackageloaded{txfonts}\@tempswafalse\@tempswatrue
\if@tempswa
  \DeclareFontFamily{U}{txsymbols}{}
  \DeclareFontFamily{U}{txAMSb}{}
  \DeclareSymbolFont{txsymbols}{OMS}{txsy}{m}{n}
  \SetSymbolFont{txsymbols}{bold}{OMS}{txsy}{bx}{n}
  \DeclareFontSubstitution{OMS}{txsy}{m}{n}
  \DeclareSymbolFont{txAMSb}{U}{txsyb}{m}{n}
  \SetSymbolFont{txAMSb}{bold}{U}{txsyb}{bx}{n}
  \DeclareFontSubstitution{U}{txsyb}{m}{n}
  \DeclareMathSymbol{\aleph}{\mathord}{txsymbols}{64}
  \DeclareMathSymbol{\beth}{\mathord}{txAMSb}{105}
  \DeclareMathSymbol{\gimel}{\mathord}{txAMSb}{106}
  \DeclareMathSymbol{\daleth}{\mathord}{txAMSb}{107}
\fi
\makeatother

\def\authorone{Peter T. Breuer}

\def\E{{\cal{E}}}
\def\enc[#1]{\E[#1]}
\def\D{{\cal{D}}}
\def\dec[#1]{\D[#1]}
\def\QED{\raisebox{0.8ex}{\framebox{\kern0.2ex}}}
\def\C#1{{\mathbb C}[#1]}
\def\S#1{{\mathbb C}^L[#1]}

\def\H{{\rm H}}

\DeclareFontShape{OT1}{cmtt}{bx}{n}
     {
      <5> <6> <7> <8> <9>
      <10> <10.95> <12> <14.4> <17.28> <20.74> <24.88> cmbtt10
      }{}
\DeclareFontShape{OT1}{cmtt}{b}{n}
  {<->sub * cmtt/bx/n}{}

\newtheorem{lemma}{Lemma}
\newtheorem{proposition}{Proposition}
\newtheorem{corollary}{Corollary}
\newtheorem{theorem}{Theorem}

\newtheorem{definition}{Definition}

\makeatletter
\def\blfootnote{\xdef\@thefnmark{}\@footnotetext}
\makeatother

\makeatletter
\DeclareRobustCommand*\cal{\@fontswitch\relax\mathcal}
\makeatother

\DeclareFloatingEnvironment[fileext=box,placement={!tb},name=Box]{mybox}

\newenvironment{frameenv}[1][]
{%
 \begin{mybox}[#1]%
 \begin{framed}
 \begin{minipage}{0.99\columnwidth}
 \begin{footnotesize}
}{%
 \end{footnotesize}\end{minipage}\end{framed}\end{mybox}%
}

\newcounter{myfootnote}
\newcommand{\myfootnote}[2]{%
\let\oldthefootnote=\thefootnote%
\renewcommand{\thefootnote}{#1}%
\footnote{#2}%
\addtocounter{myfootnote}{1}%
\let\thefootnote=\oldthefootnote%
}

\begin{document}
\pagestyle{headings}  

\title{
\hbox to\textwidth{\hss Compiled  Obfuscation for Data Structures\hss}
in Encrypted Computing
}

\author{
\IEEEauthorblockN{\authorone}
\IEEEauthorblockA{\textit{Hecusys LLC}\\
                  \textit{Atlanta, GA, USA}\\
                  ptb@hecusys.com
                 }
}

\maketitle

\begin{abstract}
Encrypted computing is an emerging technology based on a processor that
`works encrypted', taking encrypted inputs to encrypted outputs while
data remains in encrypted form throughout.  It aims to secure user data
against possible insider attacks by the operator and operating system
(who do not know the user's encryption key and cannot access it in the
processor).  Formally `obfuscating' compilation for encrypted computing is
such that on each recompilation of the
source code, machine code of the same structure is emitted for which
runtime traces also all have the same structure but each word beneath
the encryption differs from nominal with maximal possible entropy
across recompilations.  That
generates classic {\em cryptographic semantic security} for 
data, relative to the
security of the encryption,  but it guarantees only single words
and an adversary has more than that on which to base decryption
attempts.  This paper extends the existing integer-based technology to
doubles, floats, arrays, structs and unions as data structures,
covering {\sc ansi} C. A single principle drives compiler design 
and improves the existing security theory to quantitative results:
every arithmetic instruction that writes must vary to the maximal extent
possible.
\end{abstract}

\begin{IEEEkeywords}
obfuscation, compilation, encrypted computing
\end{IEEEkeywords}

\section{{Introduction}}
\label{s:Intro}

\noindent
This article examines how to make `formally obfuscating'
compilation for
encrypted computing work for the complex data structures of standard
programming languages such as {\sc ansi} C \cite{ansi99}, with its long long,
float, double, array, struct (record) and union data types.  How to do it for
32-bit integer-only computing was established in \cite{BB13a} (and is
recapitulated in Section~\ref{s:Comp}).
Integers are enough for formal purposes but this paper bootstraps that
to cover practice and heterogeneous data structures with a simple
approach that reworks all the theory.

Encrypted computing means running on a processor that `works profoundly
encrypted' in user mode (in which access is always limited to certain
registers and memory), taking encrypted inputs to encrypted outputs via
encrypted intermediate values in registers and memory.  The processor
works unencrypted in the conventional way in operator mode, which has
unrestricted access to all registers and memory.  Since user data exists
only in encrypted form, operator-level privilege gives no `magic'
access to the decrypted form of user data (the user can interpret the
data -- elsewhere -- as they know the key).  Several prototype
processors that support encrypted computing at near conventional speeds
already exist (see Section~\ref{s:Back}).

Platform issues such
as the real randomness of random numbers or power side-channel
information leaks are not at question here.
Keys may be installed at manufacture, as with Smartcards
\cite{SmartCard}, or uploaded in public view to the \mbox{write-only}
internal store via a Diffie-Hellman circuit \cite{buer2006cmos}, and are
not accessible via the programming instruction interface.
Key management is not an issue via a simple argument:
if (a) user B's key is still loaded when user A runs, then A's programs
do not run correctly because the running encryption is wrong for them,
and if (b) B's key is in the machine together with B's program when A
runs, then user A cannot supply appropriate encrypted inputs nor
interpret the encrypted output.  The question of security user on user
essentially boils down to security for user mode against operator mode
as the most powerful potential adversary, and it is proved in
\cite{BB18c} that (i) a processor that supports encrypted computing,
(ii) an appropriate machine code instruction set architecture, (iii) a
compiler with an `obfuscating' property, together formally provide
classic cryptographic semantic security \cite{Goldwasser1982} (CSS),
relative to the security of the encryption, for user data against
operator mode as adversary.  A translation is that encrypted computing
cannot in itself further compromise the encryption, and `good' security
amounts to choosing secure encryption.

The obfuscation property (iii) for the compiler
simply requires it to produce code such that an adversary cannot count
on 0,\,1,\,2 and other low values being the most common to appear
(encrypted) in a program trace.  That would be the case if the program
were written by a human and compiled to machine code conventionally, and
it would allow statistically-based dictionary attacks \cite{katz1996}
against the encryption.  The property is that no value may
appear with any higher frequency than any other, both for
observations of single words and for simultaneous observations at
multiple points in a trace. The property is violated, for
example, in implementations \cite{DGHV10} of fully homomorphic encryptions
\cite{RAD78,Gentry09} (FHE), where the output of a 1-bit AND (multiplication)
operation is predictably 0 with 75\% probability (see
Box~\ref{b:1}a).\footnote{That 0 is 
a probable outcome from multiplication in a FHE $\E$ is not an extra liability
because in 1-bit arithmetic $\E[x]+\E[x]=\E[0]$ with certainty
from any observed encrypted value $\E[x]$.  It can also be relied on
that $\E[1]$ is one of the inputs in any nontrivial calculation because
`all-zeros' as inputs propagates through to all-zeros as output via
$\E[0]+\E[0]=\E[0]*\E[0]=\E[0]$.  }

\begin{frameenv}[t]
\begin{flushleft}
\small
\refstepcounter{mybox}
\centerline{\rm Box \arabic{mybox}}
\medskip
\begin{minipage}[t]{0.48\columnwidth}
{\footnotesize
(a) A fully homomorphic encryption (FHE) $\E$ of 1-bit data
does not have the cryptographic semantic security (CSS) property.}
\label{b:1}
\begin{align*}
\E[0]*\E[0]&=\E[0]\\
\E[0]*\E[1]&=\E[0]\\
\E[1]*\E[0]&=\E[0]\\
\E[1]*\E[1]&=\E[1]
\end{align*}
\noindent
Guessing 0 as outcome is right 75\% of the time.
\end{minipage}
\hfill
\begin{minipage}[t]{0.48\columnwidth}
(b) A FHE program that adds 2-bit data to itself:

\bigskip

\begin{align*}
\E[0]+\E[0]&=\E[0]\\
\E[1]+\E[1]&=\E[2]\\
\E[2]+\E[2]&=\E[0]\\
\E[3]+\E[3]&=\E[2]
\end{align*}
has output $y=2x$ that is 100\% even, breaking CSS.
\end{minipage}
\end{flushleft}
\end{frameenv}

This document will use `the operator' for operator mode.  A subverted
operating system is `the operator', as is a human with administrative
privileges, perhaps obtained by physically interfering with the boot
process.  A scenario for an attack by the operator is where
cinematographic imagery is being rendered in a server farm.  The
computer operators have an opportunity to pirate for profit portions of
the movie before release and they may be tempted.  Another 
scenario is the processing in a specialised facility of satellite photos
of a foreign power's military installations to spot changes.
If an operator (or hacked operating system) can modify
the data to show no change where there has been some, then that is an
option for espionage.  A {\em successful attack} by the operator is one
that discovers the plaintext of user data or alters it to order.

A processor starts in operator mode when it is switched on, in order to
load operating system code into reserved areas of memory from disk, and
conventional application software relies on the processor to change from
user mode to operator mode and back for the operating system system
support routines (e.g., disk I/O) as required, so the operator mode of
working of the processor intrinsically presents difficulties as an
adversary.  Nevertheless, the CSS result of \cite{BB18c} means the
operator cannot directly or indirect\-ly by deterministic or stochastic
means read a word of user data beneath the encryption, even to a
probability slightly above chance.  Nor can user data be rewritten
deliberately, even stochas\-tically on the balance of averages, to a
value beneath the encryption that is indep\-e\-n\-dently defined, such
as $\pi$, or the encryption key (see {\cite{BB18c}} again).
That is a good start on answering (positively) to the question of the
security of encrypted computing as a whole, but it might be, for
example, that an adversary can detect that an anomaly in satellite photos
has been found, though they cannot tell what it is.  A simple example
is a one-instruction program that adds its input to itself
(Box~\ref{b:1}b).  An observer would not know what the input is nor what
the output is, but can be sure that the latter is twice the former.  It
terms of pairs $x$, $y$ of input/output values, only four of sixteen
are possible, making a statistical dictionary attack feasible.  Ideally,
a compiler for encrypted computing should produce program codes such
that biases in joint frequencies of values beneath the encryption are
removed.  In principle, it can do that by injecting some very noisy signal
of its own that swamps any existing biases.

An `obfuscating' compiler (iii) like that is described in \cite{BB17a}
and is proved
in \cite{BB18c} to generate object code that varies on recompilation of
the same source code but always looks the same to an adversary, the
difference consisting entirely of the encrypted constants embedded in the
code (which the adversary a priori cannot read, lacking the encryption
key).  Runtime traces also `look the same,' with the same instructions n
the same order, the same jumps and branches, reading from and writing to
the same registers.  But the data beneath the
encryption varies arbitrarily and independently from recompilation to
recompilation at each point in the trace, subject only to the
constraints that a copy instruction preserves the value, and the
variations introduced by the compiler are always equal where control
paths join (i.e., at either end of a loop, after conditional blocks, at
subroutine returns, at either end of a goto). Within those constraints,
compiled codes vary as much as is possible, in a way that can be
quantified precisely. A new principle subsuming that 
is put forward here:
\begin{equation}
\parbox{0.85\columnwidth}{
\em Every arithmetic instruction that writes should
introduce maximal possible entropy to the program trace.}
\tag{$\widetilde{\mathds{H}}$}
\label{e:maxH}
\end{equation}
as a single driver for the approach, reworking existing theory.

Entropy is measured across recompilations, so what this means is that
the compiler fully exercises its possibilities for varying the trace at
each opportunity in a compiled program.  It does not, for example,
always use 1 as the increment in an addition instruction when the
possibility exists of doing something different.  If two addition
instructions are introduced, then both vary independently across
compilations.
The principle \eqref{e:maxH} allows CSS and stronger
formulations of security relative to the security of encryption to be
proved (see Section~\ref{s:Theory}).

The compiler of \cite{BB17a} implements the principle \eqref{e:maxH} for a
minimal C-like language with 32-bit signed integers beneath the
encryption as the only data type.  The extension of compilation to {\sc
ansi} C pointers, arrays, structs (record types) and unions, arbitrarily
nested, will be described in this paper.  All atomic data types (int,
short int, long int, long long int, signed and unsigned, float and
double float) are covered.  Pointers must be declared as restricted to a
named area of memory (an array), which is a limitation with respect to
the standard.

Encrypted 32-bit integer arithmetic will be taken as primitive.
Since hardware is not the focus here, for further
convenience, encrypted 64-bit integer arithmetic will also be assumed
for the target platform, carried out on two encrypted 32-bit integers
representing the high and low bits respectively (that can be
supported in software, as an alternative).

Encrypted 32-bit floating point arithmetic will also be taken as
primitive, on the same rationale. It works on encrypted
32-bit integers each encoding a 32-bit float bitwise as specified in
IEEE standard 754 (ISO standard 60559; see the good commentaries on the
standard in \cite{cody1981analysis} and \cite{Goldberg1991}).  
Encrypted 64-bit floating point arithmetic will be taken as primitive
too, working on two encrypted 32-bit integers encoding separately
the high and low bits of a 64-bit double float as per the IEEE 754
standard.  All these primitives are supported by at least one of the
prototype processors referenced in Section~\ref{s:Back}.
Coincidentally, the IEEE floating point test suite at
\url{http://jhauser.us/arithmetic/TestFloat.html} consisting of 22,000
lines of C code is one of the compilation and execution tests for our
own prototype compiler, so we can be sure that encrypted floating point
arithmetic in software would work if we had to resort to it, and that our
test platform's implementation in hardware is correct.

This article is organised as follows.
%
Section~\ref{s:Back} introduces extant platforms for encrypted
computing and discusses known elements of the theory.
Section~\ref{s:FxA} introduces a modified OpenRISC
(\url{http://openrisc.io}) machine code instruction set for encrypted
computing first described in \cite{BB17a}, and its
abstract semantics.
Section~\ref{s:Comp} resumes `obfuscating' integer-based compilation
as in \cite{BB17a}. Section~\ref{s:Long} extends it to ramified
basic types such as long integers and floats, Section~\ref{s:Arr} deals
with arrays, Section~\ref{s:Struct} with `struct' (record) types, and
Section~\ref{s:Union} with union types. The theory is developed in
Section~\ref{s:Theory}, quantifying the entropy
in a runtime trace for code compiled according to the principle
\eqref{e:maxH} and characterising the compilation as `best possible'
with respect to that.  Section~\ref{s:Discuss} discusses the further
implications for security in this context.

\section*{{Notation}}
\label{s:Not}

\noindent
Encryption is denoted by $x^\E=\E[x]$ of plaintext value $x$.
Decryption is $x=\D[x^\E]$. The operation on the
ciphertext domain corresponding to $f$ on the plaintext domain is
written $[o]$, where $x^\E\mathop{[o]}y^\E=\E[x\mathop{o} y]$.

\section{Background}
\label{s:Back}

Several fast processors for encrypted computing are described in
\cite{BB18b}.  Those include the 32-bit KPU \cite{BB16b} with 128-bit
AES encryption \cite{DR2002}, which benchmarks at approximately the speed of
a 433\,MHz classic Pentium, and the slightly older 16-bit HEROIC
\cite{oic,heroic} with 2048-bit Paillier encryption \cite{Pail99}, which
runs like a 25\,KHz Pentium, as well as the recently announced
CryptoBlaze \cite{cryptoblaze18}, 10$\times$ faster.

The machine code instruction set defining the programming interface
is important because a conventional
instruction set is insecure against powerful insiders, who may, for
example,  steal an
(encrypted) user datum $x$ and put it through the machine's division
instruction to get $x/x$ encrypted, an encrypted 1.  Then any desired
encrypted $y$ may be constructed by repeatedly applying the machine's
addition instruction.  By using the instruction set's comparator
instructions (testing $2^{31}{\le}z$, $2^{30}{\le}z$, \dots) on an
encrypted $z$ and subtracting on branch, $z$ may be obtained
efficiently bitwise.  That is a {\em chosen instruction attack} (CIA)
\cite{Rass2016}.
The instruction set has to 
resist such attacks, but the compiler must be involved too, else
there would be {\em known
plaintext attacks} (KPAs) \cite{Biryukov2011} based on the idea that
not only do instructions like $x{-}x$ predictably
favour one value over others (the result
there is always $x{-}x{=}0$), but human programmers intrinsically use values
like 0,\,1 more often. The compiler's job is to
even out those statistics.

A compiler must do that even for object code consisting of a single
instruction.
That gives the conditions on the machine code instruction
design (first described in \cite{BB17a}) shown in Box\,\ref{tb:ins}):
instructions must (1) execute atomically, or recent attacks such as
Meltdown \cite{Lipp2018meltdown} and Spectre \cite{Kocher2018spectre}
against Intel might become feasible, must (2) work with encrypted values
or an adversary could read them, and (3) must be adjustable via embedded
encrypted constants to offset the values beneath the encryption by 
arbitrary deltas.  The condition (4) is for the security proofs and
amounts to different padding or blinding factors for encrypted program
constants and runtime values.
\begin{frameenv}[tb]
\begin{flushleft}
{\small
\refstepcounter{mybox}
\rm Box \arabic{mybox}:
Machine code conditions. Instructions \dots 
}
\label{tb:ins}
\end{flushleft}
\normalsize\em%
\begin{minipage}{0.99\textwidth}
\begin{align}
\begin{minipage}[b]{0.9\columnwidth}\small
\begin{itemize}[leftmargin=*]
\item
are a black box from the perspective of the programming
interface, with no intermediate states;
\end{itemize}
\end{minipage}
\tag{1}\\
\begin{minipage}[b]{0.9\columnwidth}\small
\begin{itemize}[leftmargin=*]
\item
take encrypted inputs to encrypted outputs;
\end{itemize}
\end{minipage}
\tag{2}\\
\begin{minipage}[b]{0.9\columnwidth}\small
\begin{itemize}[leftmargin=*]
\item
are adjustable via (encrypted) embedded constants 
to produce any offset in (decrypted) inputs/outputs;
\end{itemize}
\end{minipage}
\tag{3}\\
\begin{minipage}[b]{0.9\columnwidth}\small
\begin{itemize}[leftmargin=*]
\item
are such that there are no collisions between 
encrypted instruction constants and runtime values.
\end{itemize}
\end{minipage}
\tag{4}
\end{align}
\end{minipage}
\setcounter{equation}{4}
\end{frameenv}

In this document (4) will be strengthened to also:
\begin{equation}
\begin{minipage}[b]{0.87\columnwidth}\em%
There are no collisions between (encrypted) constants in instructions
with different opcodes, or differently positioned constants
where the opcode is equal.
\end{minipage}
\tag{4$^*$}
\label{e:4*}
\end{equation}
Padding beneath the encryption enforces that.  The aim is that
experiments by the adversarial operator that transplant constants from
one instruction to another cannot be performed. With (4),
experiments  that use
runtime encrypted data value as an instruction constant, or vice versa,
are ruled out.
With (4*) an adversary can modify copied instructions even less.

\makeatletter
\newcommand{\customlabel}[2]{%
\protected@write\@auxout{}{\string\newlabel{#1}{{#2}{\thepage}}}%
}
\makeatother

The salient effect of a machine code instruction set satisfying (1-4) is proved
in \cite{BB18c} to be:
\begin{equation}
\parbox[b]{0.87\columnwidth}{\em A machine code instruction
program and its runtime trace (with encrypted data) can be interpreted
arbitrarily with respect to the plaintext data beneath the encryption at
any point in memory and in the control graph by any observer and
experimenter who does not have the key to the encryption,
with the proviso that copy instructions preserve value and the delta from
nominal at start and end of a loop is the same.}
\tag{$\AT$}
\label{e:star}
\end{equation}
That means that picking any one point in the trace, the word beneath the
encryption there varies over a 32-bit range from recompilation to
recompilation with flat probability, independently of (almost) any other
point in the trace.  The exceptional points that are {\em not}
independent are data pairs that are the inputs to and outputs from a
copy instruction, and also, data measured in the same register or memory
location respectively at the beginning and end of a loop must have the
same deltas from nominal values beneath the encryption, whatever that
delta is. To keep programs working correctly the compiler has to
arrange that they are same.  The proviso actually holds wherever two
control paths join in the machine code, at the beginning of a loop but
also at the target of any jump or conditional branch, in particular at
the label of a backward-going jump and multiple entry or exit points of
a subroutine.

The rationale behind \eqref{e:star} is that an arbitrary delta from the
nominal value can be introduced by the compiler in one instruction and
changed again in the next instruction, via the embedded instruction
constants of (3), while (1-2) prevent the adversary from knowing.
Note that (1) means `no side-channels'.
The compiler's job boils down to:
\begin{equation}
\parbox[b]{0.85\columnwidth}{\em Varying the encrypted instruction
constants \rm (3) \em from recompilation to recompilation so deltas
from nominal in the runtime data beneath the encryption at each point in
the trace are equiprobable.}
\tag{$\flat$}
\label{e:S}
\end{equation}
The compiler strategy in \cite{BB17a} does that. It
is subsumed by \eqref{e:maxH} here, but
\cite{BB18c} shows \eqref{e:star}  implies \eqref{e:S}, which in turn implies:
\begin{equation}
\parbox[b]{0.85\columnwidth}{\rm Cryptographic semantic
security (CSS) \em holds for user data against insiders not privy to the
encryption.}
\tag{$\padlock$}
\label{e:ddagger}
\end{equation}
I.e., {\em encrypted computation does not compromise encryption}.

How the `equiprobable variation' is obtained
by the compiler is encapsulated in Box \ref{tb:how}:
a new {\em ob\-fus\-cation scheme} is generated at each
recompilation. That is
a planned {\em offset delta for the data beneath the encryption in
every memory and register location per point in the program
control graph}.

Precisely, the compiler $\C{-}$ translates an expression $e$ that
is to end up in register $r$ at runtime into machine code $\it mc$
and generates a 32-bit offset $\Delta e$ for $r$ at the point in the
program where
it is loaded with the result of the expression $e$. That is
\begin{equation}
\C{e}^r = ({\it mc},\Delta e)
\end{equation}
The offset $\Delta e$ is the obfuscation for
register $r$ at the point where the encrypted value of
the expression is written to it.

Let $s(r)$ be the content of register $r$ in state $s$ of the processor
at runtime. The machine code {\em mc}'s action 
changes state $s_0$ to an $s_1$ with a ciphertext in $r$ whose plaintext
value differs by $\Delta e$ from the nominal value $e$:
\begin{equation}
s_0 \mathop\rightsquigarrow\limits^{\it mc}  s_1 ~\text{where}~ s_1(r) = 
\E[e + \Delta e]
\end{equation}
Bitwise exclusive-or or the binary operation of another
mathematical group are alternatives to addition in the $e + \Delta e$.

The encryption $\E$  is shared with the user and the processor but
not the potential adversaries: the operator and operating system. The
randomly generated offsets
$\Delta e$ of the obfuscation scheme are known to the user, but not
the processor and not the
operator and operating system. The user compiles the program and sends
it to the processor to be executed and needs to know the
offsets on the inputs and outputs.  That allows the right
inputs to be created and sent off for processing on the encrypted
computing platform, and allows sense to be made of the outputs received
back.

\begin{frameenv}[t]
\begin{flushleft}
\small
\refstepcounter{mybox}
\rm Box \arabic{mybox}:
\footnotesize
What the compiler does:
\label{tb:how}
\end{flushleft}
\begin{enumerate}
\item[(A)] {\sl change only encrypted program constants, generating
via (3) an {\rm obfuscation scheme} of planned} {\rm offsets from nominal values} {\sl for 
instruction inputs and outputs beneath the encryption};
\end{enumerate}
\begin{enumerate}
\item[(B)] {\sl make runtime traces look unchanged, apart
from differences in the (encrypted) instruction constants and data}
(A);
\end{enumerate}
\begin{enumerate}
\item[(C)] {\sl
equiprobably\,generate\,all\,obfuscation\,schemes\,satisfying}\,(A),\,(B).
\end{enumerate}
\end{frameenv}

\section{{FxA Instructions}}
\label{s:FxA}

\noindent
A `fused anything and add' (FxA) 
\cite{BB17a} instruction set architecture (ISA) is the general target here,
satisfying
conditions (1-4) of Section~\ref{s:Intro}.  The integer
portion is shown in Table~\ref{tb:1}. It is adapted from the open standard
OpenRISC instruction set v1.1 \url{http://openrisc.io/or1k.html}. That
has about 200 instructions (6-bit opcode plus variable minor
opcodes) separated into single and double precision
integer and floating point and
vector subsets with instructions all 32 bits long and the FxA
instruction set follows that design closely. FxA instructions, like
OpenRISC instructions, access
up to three 32 general purpose registers (GPRs) per instruction,
designated in
contiguous 5-bit plaintext specifier fields within the instruction.
\begin{table}[tbp]
\caption{Integer portion of FxA machine code instruction set for
encrypted working --  abstract syntax and semantics.}
\label{tb:1}
{\centering
\scriptsize
\begin{tabular}{@{}l@{\,}c@{\kern1pt}c@{\kern1pt}l@{\kern1pt}l@{~}l@{}}
\em op.&
    \multicolumn{4}{@{}l}{\em fields} 
     & \multicolumn{1}{@{}l}{\em \hbox to 0.39in {mnem.\hfill} semantics \hfill}\\
\hline\\[-1ex]
add &$\rm r_0$
  &$\rm r_1$
    &$\rm r_2$
      &\kern0pt$k^\E$
        &\hbox to 0.39in {add} $r_0{\leftarrow}
           r_1\kern0.5pt\mathop{[+]} r_2\mathop{[+]} k^\E$\\
sub &$\rm r_0$
  &$\rm r_1$
    &$\rm r_2$
      &\kern0pt$k^\E$
        &\hbox to 0.39in {subtract} $r_0{\leftarrow}
           r_1\kern0.5pt\mathop{[-]} r_2\mathop{[+]} k^\E$\\
mul &$\rm r_0$
  &$\rm r_1$
    &$\rm r_2$
      &\kern0pt$k^\E_0\kern0pt k^\E_1\kern0pt k^\E_2$
       &\hbox to 0.39in {multiply}
       $r_0{\leftarrow}(r_1\kern0pt\mathop{[-]}k^\E_1\kern0pt)\mathop{[\,*\,]}(r_2\kern0pt\mathop{[-]}k^\E_2\kern0pt)\mathop{[+]}k^\E_0$\\[0.5ex]
div &$\rm r_0$
  &$\rm r_1$
    &$\rm r_2$
      &\kern0pt$k^\E_0\kern0pt k^\E_1\kern0pt k^\E_2$
       &\hbox to 0.39in {divide} 
       $r_0{\leftarrow}(r_1\kern0pt\mathop{[-]}k^\E_1\kern0pt)\mathop{[\div]}(r_2\kern0pt\mathop{[-]}k^\E_2\kern0pt)\mathop{[+]}k^\E_0$\\
\dots\\
mov&$\rm r_0$&$\rm r_1$& &   
     &\hbox to 0.39in {move}
       $r_0{\leftarrow} r_1$\\
beq& $i$&$\rm r_1$&$\rm r_2$&$k^\E$  
     &\hbox to 0.39in {branch}
       ${\rm if}\,$b$\,{\rm then}\,{\it pc}{\leftarrow}{\it pc}{+}i$,
       $b\Leftrightarrow r_1 \mathop{[=]} r_2 \mathop{[+]} k^\E$\\
bne& $i$&$\rm r_1$&$\rm r_2$&$k^\E$  
     &\hbox to 0.39in {branch}
       ${\rm if}\,$b$\,{\rm then}\,{\it pc}{\leftarrow}{\it pc}{+}i$,
       $b\Leftrightarrow r_1 \mathop{[\ne]} r_2 \mathop{[+]} k^\E$\\
blt& $i$&$\rm r_1$&$\rm r_2$&$k^\E$   
     &\hbox to 0.39in {branch}
       ${\rm if}\,$b$\,{\rm then}\,{\it pc}{\leftarrow}{\it pc}{+}i$,
       $b\Leftrightarrow r_1 \mathop{[<]} r_2 \mathop{[+]} k^\E$\\
bgt& $i$&$\rm r_1$&$\rm r_2$&$k^\E$   
     &\hbox to 0.39in {branch}
       ${\rm if}\,$b$\,{\rm then}\,{\it pc}{\leftarrow}{\it pc}{+}i$,
       $b\Leftrightarrow r_1 \mathop{[>]} r_2 \mathop{[+]} k^\E$\\
ble& $i$&$\rm r_1$&$\rm r_2$&$k^\E$   
     &\hbox to 0.39in {branch}
       ${\rm if}\,$b$\,{\rm then}\,{\it pc}{\leftarrow}{\it pc}{+}i$,
       $b\Leftrightarrow r_1 \mathop{[\le]} r_2 \mathop{[+]} k^\E$\\
bge& $i$&$\rm r_1$&$\rm r_2$&$k^\E$   
     &\hbox to 0.39in {branch}
       ${\rm if}\,$b$\,{\rm then}\,{\it pc}{\leftarrow}{\it pc}{+}i$,
       $b\Leftrightarrow r_1 \mathop{[\ge]} r_2 \mathop{[+]} k^\E$\\
\dots \\
b   &$i$&   &   &   
     &\hbox to 0.39in {branch}
       ${\it pc}\leftarrow {\it pc}+i$\\
sw   &\multicolumn{4}{@{}l@{}}{$(k_0^\E){\rm r_0}~{\rm r_1}$}
     &\hbox to 0.39in {store}
       $\mbox{\rm mem}\llbracket r_0\mathop{[+]}k_0^\E\rrbracket \leftarrow r_1$ \\
lw   &\multicolumn{4}{@{}l@{}}{${\rm r_0}~(k_1^\E){\rm r_1}$}
     &\hbox to 0.39in {load}
       $r_0 \leftarrow \mbox{\rm mem}\llbracket r_1\mathop{[+]}k_1^\E\rrbracket
              $\\
jr   &$\rm r$&  &    &
     &\hbox to 0.39in {jump}
       ${\it pc} \leftarrow r$\\
jal  &$j$&      &    &
     &\hbox to 0.39in {jump}
       ${\it ra} \leftarrow {\it pc}+4;~{\it pc} \leftarrow j$\\
j   &$j$&      &    &
     &\hbox to 0.39in {jump}
       ${\it pc} \leftarrow j$\\
nop &   &      &    &
     &\hbox to 0.39in {no-op}\\[2ex]
\end{tabular}
}
\noindent
\begin{minipage}{0.99\columnwidth}
{\scriptsize\sc Legend}\\[0.5ex]
\scriptsize
\noindent
\begin{tabular}{@{}l@{~}c@{~}l@{\quad}l@{~}c@{~}l@{\quad}l@{~}c@{~}l@{}}
r &--&register indices
 & $k$&--&32-bit integers
 & pc &--&prog.\ count\ reg.\\
$j$&--&program count
 & `$\leftarrow$'&--&assignment
 & ra &--&return addr.\ reg.\\
$\enc[\,~\,]$&--&encryption
 & $i$&--&pc increment
 & $r$ &--& register content \kern-5pt \\
$k^\E$&--&encrypted value $\enc[k]$
 &  \multicolumn{3}{@{}l@{}}{$x^\E\mathop{[o]}y^\E =
         \E[x\mathop{o}y]$}\kern-20pt
 &  \multicolumn{3}{@{}l@{}}{$x^\E\mathop{[R]}y^\E \Leftrightarrow
         x\mathop{R}y$}\kern-20pt
\end{tabular}
\end{minipage}
\end{table}

To give an idea of what FxA machine code looks like `in action', a trace
of code compiled for the Ackermann function\footnote{Ackermann C code: {\bf int}
A({\bf int} m,{\bf int} n) \{ {\bf if} (m == 0) {\bf return} n+1; {\bf
if} (n == 0) {\bf return} A(m-1, 1); {\bf return} A(m-1, A(m, n-1));
\}.} \cite{Sundblad71} is shown in Table~\ref{tab:3}.
For readability here, the final delta for the return value in register
{\bf v0} is set to zero.  That function is the most computationally
complex function theoretically possible, stepping up in complexity for
each increment of the first argument, so it is a good test of 
correct compilation.

\subsection{Prefix Instructions}
\noindent
FxA instructions may need to contain 128-bit or longer encrypted
constants, so some adaptation of the basic OpenRISC architecture is
required for that to be possible.  A `prefix' instruction takes care of
it, supplying extra bits as necessary.  Each prefix instruction instruction s 32
bits long, but several may be concatenated.

\def\stardot{\mathop{*}\limits^.}

\subsection{Single Precision Floating Point}
\noindent
In addition to the integer instructions of Table~\ref{tb:1}, 
there may be floating
point instructions {\bf addf}, {\bf subf}, {\bf mulf} etc.\ paralleling the
OpenRISC floating point subset.  The contents of
registers and memory for floating point operations are the
encryptions of 32-bit integers that themselves
encode floating point numbers (21 mantissa bits, 10 exponent
bits, 1 sign bit) via the IEEE\,754 standard encoding.

\begin{definition}
Let $\stardot$ denote the floating point multiplication on
plaintext integers encoding IEEE~754 floats, and use the same convention
for other arithmetic operations and relations.
\end{definition}
\noindent
Let
$[\,\stardot\,]$ be the corresponding operation in the ciphertext domain,
following the notation convention at end of Section~\ref{s:Not}.
Then the floating point multiplication instruction semantics is
\begin{equation}
\tag{$\stardot$}
\label{e:dagger*f}
r_0 \leftarrow (r_1 \mathop{[-]} k^\E_1) \,\mathop{[\,\stardot\,]}\, (r_2 \mathop{[-]} k^\E_2) \mathop{[+]} k^\E_0
\end{equation}
The $-$ and $+$ are the ordinary
plaintext integer subtraction and addition operations respectively,
and $[-]$ and $[+]$ are the corresponding operations in the
ciphertext domain (see Notation in Section~\ref{s:Not}). That is,
the FxA floating point multiplication takes the encrypted integers
representing (in IEEE 754 format) floating point numbers that have been
offset as integers, undoes the offsets then multiplies them as floats,
obtaining the IEEE 754 integer representation
before offsetting as integer again.  The operation is atomic, as
required by (1) of Box~\ref{tb:ins}, leaving no trace if aborted.
The offsets $k_i$ satisfy the requirement (3).

\def\eqdot{\mathop{=}\limits^.}

The FxA set in use in our prototypes 
has two encrypted constants for a floating point test condition in
branch instructions.  The floating point branch-if-equal
instruction calculates 
\begin{equation}
\tag{$\eqdot$}
\label{e:dagger=f}
(r_1\mathop{[-]}k_1^\E) \,\mathop{[\eqdot]}\, (r_2\mathop{[-]} k_2^\E)
\end{equation}
where $\eqdot$  is the floating point comparison on integers encoding
floats via IEEE\,754, and $[\eqdot]$ is the corresponding
test in the ciphertext domain, with
$x^\E\mathop{[\eqdot]}y^\E \Leftrightarrow x\eqdot y$.  The subtraction
is as integers on the encoding, not floating point. The operation
is atomic, leaving no trace if aborted or interrupted, as required
by (1) of Box~\ref{tb:ins}, and all encrypted operations in the processor
(should and do) take the same time and power on all operands.

\subsection{Instruction Diddling}

\noindent
Condition (2) of Box~\ref{tb:ins} requires there to be one more constant
physically present in each branch instruction, an encrypted bit $k_0$
that decides if the 1-bit result of the test is to be inverted or not.
That is because the test outcome is observable by whether the branch is
taken or not, so by condition (3) it should be variable via an encrypted
constant in the instruction.  The bit changes equals to not-equals
and vice versa, a less-than into a greater-than-or-equal-to, and so on.
The bit is said to {\em diddle} the instruction.  In practice, the bit
is composed from the padding bits in the other constants in the
instruction, so it has not been mentioned explicitly in
Table~\ref{tb:1}, where the branch semantics shown are after the diddle.

The opcode in the instruction is in plaintext, but which branch 
in the control graph is which is hidden by the diddle.

\subsection{The Debatable Equals Branch Instruction}
\label{ss:Debate}

\noindent
Diddling works well to disguise less-than instructions and order
inequalities in general, but not for equals versus not-equals.  What the
instruction is, equals or not-equals, may be guessed by what
proportion of operands cause a jump at runtime.  If almost all do then
that is a not-equals test.  If few do then that is an equality test.
Trying the same operand both sides is almost guaranteed to cause
equality to fail because of the embedded constants $k_1$, $k_2$ in
\eqref{e:dagger=f}, so if it succeeds instead, the equality instruction
has likely been diddled to not-equals.


So whether the test succeeds or not at runtime is detectable in practice
for an equality/not-equals branch instruction, contradicting (2).  To
beat that, the compiler described in \cite{BB17a} randomly changes the
way it interprets the original boolean source code expression at every
level so it cannot be told if the source code, not the object code, had
an equality or an not-equals test.  It independently and randomly
decides as it works upwards through a boolean expression if the source
code at that point is to be interpreted by a {\em truthteller}, who says
`true' when true is meant and `false' when false is meant, or by a {\em
liar}, who says `false' when true is meant and `true' when false is
meant.  It equiprobably generates, at each level in the boolean
expression, liar code and uses the branch-if-not-equal machine code
instruction for an equality test, or truthteller code and uses the
branch-if-equal instruction.

\def\ME{\(\E\)}
\begin{table}[!tbp]
\caption{Trace for Ackermann(3,1), result 13.}
\label{tab:3}
\centering
\begin{minipage}{0.98\columnwidth}
\scriptsize
\begin{alltt}
{\rm{PC}}  {\rm{instruction}}                           {\rm{update trace}}
\dots
35  add t0 a0  zer \ME[-86921031]       t0 = \ME[-86921028]
36  add t1 zer zer \ME[-327157853]      t1 = \ME[-327157853]
37  beq t0 t1  2   \ME[240236822]                  
38  add t0 zer zer \ME[-1242455113]     t0 = \ME[-1242455113]
39  b 1                                             
41  add t1 zer zer \ME[-1902505258]     t1 = \ME[-1902505258]
42  xor t0 t0  t1  \ME[-1734761313] \ME[1242455113] \ME[1902505258]
                                      t0 = \ME[-17347613130]
43  beq t0 zer 9   \ME[-1734761313]                 
53  add sp sp  zer \ME[800875856]       sp = \ME[1687471183] 
54  add t0 a1  zer \ME[-915514235]      t0 = \ME[-915514234] 
55  add t1 zer zer \ME[-1175411995]     t1 = \ME[-1175411995]
56  beq t0 t1  2   \ME[259897760]                   
57  add t0 zer zer \ME[11161509]        t0 = \ME[11161509]   
\dots
143 add v0 t0  zer \ME[42611675]        v0 = \ME[13]
\dots
147 jr  ra                             \# {\rm{(return \ME[13] in{\it v0})}}
\end{alltt}
\end{minipage}
\flushleft
\noindent
\small
Legend: (registers) a0 = function argument; sp = stack pointer; t0, t1 =
temporaries; v0 = return value; zer = null placeholder.
\vspace{-2ex}
\end{table}

With that compile strategy, if the equals branch instruction jumps
or not at runtime does not relate statistically to what the boolean in
the source code should be.  Condition (3) of Box~\ref{tb:ins} on the
output of the instruction is effectively vacuous with respect to the
source, as there is no definite meaning to it jumping.  An observer who
sees it jump does not know if that is the result of a truthteller's
interpretation of an equals test in the source code and it has come out
true at runtime, or it is the result of the liar's interpretation and it
has come out false.  Ditto not-equals.  This equates to a
(structured) {\em garbled circuit} construction in the classical sense
of \cite{yao86}.  While a structured boolean expression reveals its
intermediates as outputs to an observer too, the classical result
has it that no output values can be deciphered by an observer who does
not already know which is being `lied' about, and which not.

For other comparison tests, just as many operand pairs cause a branch
one way as the other,\footnote{In 2s complement arithmetic $x<y$ is the
same as $x-y=z$ and $z<0$ and exactly half of the range satisfies $z<0$
and exactly half satisfies $z\ge0$.} and make it indistinguishable
whether the opcode is diddled or not.  Still, the truthteller/liar
compile strategy is used there too.  An equality test cannot be
recreated by an adversary as $x{\le}y$ and $y{\le}x$ because only
$x{\le}y{+}k$ is available in FxA, for unknown constant $k$.  Reversing
operands is allowed by (4*) but produces $y{\le}x{+}k$, not
$y{+}k{\le}x$.  An estimate for $k$ can be made by the proportion of
pairs $(x,y)$ that satisfy the conjunction of the inequality and the
reversed inequality.  In particular whether $k{<}0$ is signalled by the
absence of pairs that satisfy both inequalities.  But diddling means the
conjunctions might be $x{>}y{+}k$ and $y{>}x+{k}$ instead, and those
have no solutions when ${-}k{-}1$ is negative.  So either $k{<}0$ or
$k{\ge}0$, which gives nothing away.

Note for the general description below of the compiler strategy
established in \cite{BB17a} that `liar' amounts to adding a delta equal
to 1 mod 2 to a boolean 1-bit result, and `truthteller' amounts to
adding a delta equal to 0 mod 2.



\section{{Obfuscating Compilation}}
\label{s:Comp}

\noindent
A compiler built to obfuscate in the sense of this article works with a
database $D : {\rm Loc}\,{\to}\,{\rm Off}$ containing a (here 32-bit)
integer offset $\Delta l$ of type Off for data in register or memory
location $l$ (type Loc).  The offset is a delta by which the runtime
data underneath the encryption is to vary from nominal at a given point
in the program, and the database $D$ comprises the {\em obfuscation
scheme}.  It is varied by the compiler as it makes a pass through the
source code.

The compiler (any compiler) also maintains a conventional database of
type $L:{\rm Var}\to {\rm Loc}$ binding source variables to registers
and memory locations.  In our prototype an intermediate layer (RALPH:
Register ALlocation in Physical Hardware) optimises the mapping and
detail of this is omitted here.

\subsection{Expressions}

In \cite{BB17a}, a generic (non-side-effecting) integer
expression compiler putting its result in register $r$
is described with type:
%
\begin{align}
\S{\_ : \_}^r &:\hbox{DB}\times\hbox{Expr} \to \hbox{MC}\times\hbox{Off}
\end{align}
where MC is the type of machine code, a sequence of FxA instructions
{\em mc}, and Off is the type of the integer offset $\Delta r$ from
nominal that the compiler intends for the result in $r$
beneath the encryption when the machine code is evaluated at runtime.
The aim is to satisfy \eqref{e:S} by varying $\Delta r$
arbitrarily and equiprobably from recompilation to recompilation.

To translate $x{+}y$, for example, where $x$ and $y$ are signed integer
expressions, the compiler first 
emits machine code $\hbox{\em mc}_1$
computing expression $x$ in register $r_1$ with offset $\Delta x$. It then 
emits machine code $\hbox{\em mc}_2$
computing expression $y$ in register $r_2$ with offset $\Delta y$. That
is
\begin{align*}
(\hbox{\em mc}_1,\Delta x) &= \S{D : x}^{r_1}\\
(\hbox{\em mc}_2,\Delta y) &= \S{D : y}^{r_2}
\end{align*}
It then decides a random offset $\Delta e$ for the whole expression
$e=x{+}y$ and emits the FxA integer addition instruction with abstract
semantics $r {\leftarrow} r_1 {[+]} r_2 {[+]} k^\E$ 
to return the result\,in\,$r$\/:
\begin{align}
\S{D: x+y}^r &= (\hbox{\em mc}_e,\Delta e)\\
\hbox{\em mc}_e &= \hbox{\em mc}_1; \hbox{\em mc}_2;
                {\bf add}~r ~r_1 ~r_2 ~k^\E\notag\\
           k  &= \Delta e - \Delta x - \Delta y\notag
\end{align}
%
%
The final offset $\Delta e$ for the runtime result in $r$ beneath the
encryption may be freely chosen, as \eqref{e:S} stipulates.


That is carrying through the global requirement for compiler
constructions \eqref{e:maxH}: the code takes the opportunity of {\em
one} new arithmetic instruction that writes, here {\bf add}, to generate
{\em one} new, independent, randomly chosen offset $\Delta e$ for the
written location $r$.  The same will be true of the compilation of the
subexpressions $x$, $y$: each arithmetic machine code instruction
emitted introduces an independent random delta in its target.

\subsection{Statements}

\noindent
Statements do not produce a result, instead they  have a side-effect.
Let Stat be the type
of statements. The statement compiler in \cite{BB17a} works not by
returning an offset, as for expressions,
but a new scheme for offsets at multiple locations:
\begin{align}
\S{\_ : \_} &:\hbox{DB}\times\hbox{Stat} \to \hbox{DB}\times\hbox{MC}
\end{align}
\noindent
Recall that a database $D$ of type DB holds the obfuscation
scheme (the offset deltas from nominal values beneath the encryption
in all locations) as the compiler works through the code, and 
consider an assignment $z{=}e$ to a source code variable $z$, which the
location database $L$ says is bound in register $r{=}Lz$.
Let a pair in the cross product DB$\times$MC be written $D:{\it mc}$ for
readability.  First code $\hbox{\em
mc}_e$ for evaluating expression $e$ in temporary register {\bf t0} at
runtime is emitted via the expression compiler as already described:
\[
(\hbox{\em mc}_e,\Delta e) = \S{D : e}^{\bf t0}
\]
Offset $\Delta e$ is generated by the expression compiler 
for the result $e$ in {\bf t0}.  A short form add instruction with
semantics
$r \leftarrow \hbox{\bf t0} \mathop{[+]} k^\E$
to change offset $\Delta e$ to a new randomly chosen offset $\Delta' r$
in register $r$ is emitted next:
\begin{align}
\S{D: z{=}e}  &= D': ~\hbox{\em mc}_e ; {\bf add}~r~\hbox{\bf t0}~k^\E\\
                k &= \Delta' r - \Delta e\notag
\end{align}
The change to the database of offsets is at index $r$.  An initial
offset $D r=\Delta r$ changes to $D' r = \Delta' r$.  The new offset has
been freely and randomly chosen by the compiler, supporting \eqref{e:S},
and the {\em one} new arithmetic machine code instruction emitted, {\bf
add}, to write the expression in the target variable incorporates {\em
one} new random delta, supporting \eqref{e:maxH}.

\section{{Long Basic Types}}
\label{s:Long}

\noindent
Double length (64-bit) plaintext  integers $x$ can be viewed as
concatenated 32-bit integers $x=x^H \mathop.  x^L$, the high and low 32
bits of $x$ respectively.  In the processor, the encryption of $x$
occupies two registers or two memory locations, containing the encrypted
values $\E[x^H]$, $\E[x^L]$ respectively.

\begin{definition}
Encryption of 64-bit integers $x$ concatenates the encryptions of 
their 32-bit high and low bit components:
\label{d:2}
\begin{align*}
x^\E = \E[x]= \E[x^H\mathop. x^L]&=\E[x^H]\mathop. \E[x^L]
\end{align*}
\end{definition}
\noindent
The FxA instructions for dealing with encrypted 64-bit values necessarily
contain (encrypted) 64-bit constants.

\subsection{Long Long Integers}

\noindent
The 64-bit integer type is known in C as `long long'.
\begin{definition}
Let $-^2$ and $+^2$ be the two-by-two independent application of
respectively 32-bit addition and 32-bit subtraction to the pairs of 32-bit
plaintext integer high-bit and low-bit components of 64-bit integers, with
similar notation for other binary operators. I.e. and e.g.:
\begin{align*}
(u_1 \mathop. l_1)\mathop{+^2\,}(u_2 \mathop. l_2) &=
              (u_1 \mathop{+} u_2) \mathop.\, (l_1 \mathop{+} l_2)
\end{align*}
\end{definition}

\begin{definition}
Let $\tilde{*}$ denote the
usual plaintext multiplication  on 64-bit `long long' integers, and
similarly for other operators.
\end{definition}
\noindent 
The FxA 64-bit multiplication operation on operands $\E[x]$, $\E[y]$
has semantics:
\begin{equation}
\E[(x \,\mathop{-^2}\, k_1) \,\tilde{*}\,
(y \,\mathop{-^2}\, k_2) \,\mathop{+^2}\, k_0]
\tag{$\tilde{*}$}
\end{equation}
where $k_0$, $k_1$, $k_2$ are 64-bit plaintext integer constants 
embedded encrypted in the instruction as $k^\E_i$, $i=0,1,2$.
Putting it in terms of the effect on
register contents, the FxA long long multiplication
instruction semantics is:
\[
r_0^H\mathop{.}r_0^L \leftarrow (r_1^H\mathop{.}r_1^L \mathop{[-^2]}k_1^\E)
                    \mathop{[\,\tilde{*}\,]}
                  (r_2^H\mathop{.}r_2^L\mathop{[-^2]}k_2^\E) \mathop{[+^2]}
                    k_0^\E
\]
For encrypted (and unencrypted) 64-bit operations the processor
partitions the register set into pairs
referred to by one name each. In those terms the semantics
is simplified to:
\[
r_0 \leftarrow (r_1 \mathop{[-^2]}k_1^\E)
                    \mathop{[\,\tilde{*}\,]}
                  (r_2\mathop{[-^2]}k_2^\E) \mathop{[+^2]}
                    k_0^\E
\]
That is written
${\bf mull}\,r_0\,r_1\,r_2\,k_0^\E\,k_1^\E\,k_2^\E$
in assembler, following the 32-bit instruction pattern.
The operation is atomic (1).

The other instructions for `long long' integer arithmetic in FxA also
match the architecture of the corresponding 32-bit integer instruction
(Table~\ref{tb:1}), with longer encrypted constants and the
`two-at-a-time' register naming convention, and an {\bf l} suffix on
the name in assembler.
Only the different opcode and the extra prefixes distinguish the long
forms `on the wire'.

The pattern for compiled code generated for long long integer
expressions and statements on the encrypted computing platform follows
exactly that for 32-bit expressions and statements but using the `{\bf
l}' instructions. Exactly one new (64-bit)
arithmetic instruction that writes is issued with each compiler
construct. 
It contains just one 64-bit (encrypted) constant that
allows the 64-bit (i.e.  $2\times32$-bit) offset delta in the target
location to be freely chosen and generated by the compiler, supporting
\eqref{e:maxH}.  The target register or memory location pair has a
different (32-bit) delta generated for each of the pair.

\subsection{Double Floats}

%

\noindent
Double precision plaintext 64-bit floats (`double') are
encoded as two (encrypted) 32-bit integers, the top and bottom bits
respectively of a 64-bit IEEE\,754 standard integer representation.
\begin{definition}
Let $\ddot{*}$ denote the plaintext double precision
floating point multiplication on the  IEEE 754 encoding of double (64-bit)
floats as 64-bit integers rendered as two 32-bit integers, and similarly for other operations and
relations.
\end{definition}
\noindent
Let $[\ddot{*}]$ be the corresponding operation in
the cipherspace domain on two pairs of encrypted 32-bit integers. Then
the FxA multiplication instruction on encrypted 64-bit double operands in
the (pairs of) registers $r_1$, $r_2$ respectively, writing to (the
pair) register $r_0$ has semantics:
\begin{align}
 r_0 \leftarrow (r_1 \mathop{[-^2]}k_1^\E)
                    \,\mathop{[\,\ddot{*}\,]}\,
                  (r_2\mathop{[-^2]}k_2^\E) \mathop{[+^2]}
                    k_0^\E
\end{align}
where $k_i^\E$, $i=0,1,2$ are encrypted 64-bit constants embedded in
the instruction. That is written 
${\bf muld}\,r_0\,r_1\,r_2\,k_0^\E\,k_1^\E\,k_2^\E$
in assembler, following the 32-bit pattern, but with a {\bf
d} suffix on the root of the mnemonic.  The operation is atomic (1).

The pattern for the compiled code emitted for double floating point
expressions and statements on the encrypted computing platform follows
exactly that for 32-bit floating point expressions and statements (which
follows the 32-bit integer pattern) but with these `{\bf d}' instructions
instead.  Exactly one new arithmetic instruction that writes is issued
per each compiler construct for expressions or a write to a
location holding a source code variable.  The instruction contains one
64-bit (encrypted) constant that allows the 64-bit (i.e.
$2\times32$-bit) offset delta in the target location to be freely chosen
and generated by the compiler, supporting \eqref{e:maxH}.

\subsection{Short Basic Types and Casts}

\noindent
Machine code instructions that act on encrypted `short' (16-bit) or
`char' (8-bit) integers are unneeded for C because short integers
are promoted to 32-bits ones at first use.

The compiler instead generates {\em cast}s following the principle
\eqref{e:maxH} (emitting any one instruction that writes entails managing
it to vary to the fullest extent possible across recompilations).
For C, the 13 basic types (signed/unsigned char, short, int, long, long
long integer, and float and double precision float, also the single bit
\_bool type) have to be inter-converted.  Here follows the 
cast for encrypted signed 32-bit `int' to encrypted
signed 16-bit `short'.  The compiler-issued code 
moves the integer 16 places left and then 16 places right again
using one  multiplication and one division (read on
for improvement):
\begin{align}
\S{D: (\hbox{short})x}^r  &= (\hbox{\em mc}_e,\Delta e)\\
                   (\hbox{\em mc}_0,\Delta x) &= \S{D: x}^r\notag\\
                   \hbox{\em mc}_e &= \hbox{\em mc}_0 ;~
                  {\bf mul}~r~r~\E[2^{16}]~\E[\Delta x]~k^\E;
                        \notag\\
                  &\hbox{\qquad\qquad}{\bf div}~r~r~\E[2^{16}]~k^\E~\E[\Delta e]
                            \notag
\end{align}
Those are short form instructions
${\bf mul}\,r_0\,r_1\,k^\E\,k_1^\E\,k_2^\E$ and
${\bf div}\,r_0\,r_1\,k^\E\,k_1^\E\,k_2^\E$ with 
semantics 
$r_0{\leftarrow}(r_1{\mathop{[-]}}k_1^\E) \mathop{[*]} k^\E  \mathop{[+]} k_0^\E$
and
$r_0{\leftarrow}(r_1{\mathop{[-]}}k_1^\E) \mathop{[\,/\,]} k^\E \mathop{[+]}
k_0^\E$.
The constants $k$, $\Delta e$ are freely chosen for these two
`arithmetic instructions that write', in support of \eqref{e:maxH}.

But (a) the compiler must avoid encryptions of $2^{16}$ always
appearing.  Instead a register $r_1$ can be loaded with the encryption
of a random number $k_1$ and then the full-form instructions of
Table~\ref{tb:1} instead of the short forms can be used, with
$r_1\mathop{[-]}k_2^\E$ in place of $\E[2^{16}]$, where
$k_2{=}k_1{-}2^{16}$.  Then the encrypted constants that appear in the
code are uniformly distributed. Also (b) the top 16 bits should be
filled randomly, but that is taken care of in the final offset
delta $\Delta e$. That the difference between $k_1$, $k_2$ for (a) is
constant at $2^{16}$ across recompilations does not help an adversary
as the processor arithmetic does not work on instruction constants (4).

Our FxA instruction set provides integer-to-float (and vice
versa) conversion primitives for the platform.  Each embeds encrypted
constants that offset inputs and outputs arbitrarily beneath the
encryption, as required by (3).  The compiler needs just one such
instruction for an integer/float cast, containing one constant allowing
one arbitrary offset beneath the encryption in the target location
to be generated, supporting \eqref{e:maxH}.

\section{{Arrays and Pointers}}
\label{s:Arr}

\noindent
There is a natural and there is an efficient way to bootstrap integer
computation to an array A of $n$ integers and both will be discussed
briefly.
The natural way is to imagine a set of variables A$_0$, A$_1$, \dots
for the entries of the array.  That allows the compiler to translate a
lookup A[i] as a compound expression
`(i$=$0){\bf?}A$_0${\bf:}(i$=$1){\bf?}A$_1${\bf:}\dots',
while a write A[i]=x can be translated to `{\bf if} (i$=$0) A$_0$=x {\bf
else} {\bf if} (i$=$1) A$_1$=x {\bf else} \dots'.  The entries get
individual offsets from nominal $\Delta{\rm A}_0$, $\Delta{\rm A}_1$,
\dots in the obfuscation scheme maintained by the compiler.

\subsection{Single Shared Array Offset}

\noindent
While the natural approach is logically correct, it makes array access 
have complexity {\bf O}($n$).  It can trivially be improved to {\bf
O}($\log_2 n$) but that is still an overhead.
%
So we have also explored an efficient approach: array A's entries
share the same offset $\Delta{\rm A}$ from their nominal value beneath
the encryption.

Then pointer-based access becomes easier to generate code for. At compile time
where in the array the pointer will point at runtime is unknown, but 
the shared offset for all array entries may be relied on. Pointers p
must be declared with the array:
\begin{center}
       {\bf restrict} A {\bf int} *p{\bf;}
\end{center}
With this approach, the compiler 
constructs the dereference $*e$ of an expression $e$ that is a pointer
into A as follows.
It first emits code \hbox{\em mc}$_e$ that evaluates the
pointer in register $r$ with a randomly generated offset $\Delta e$ beneath the
encryption:
\begin{align*}
        (\hbox{\em mc}_e,\Delta e)&= \S{D: e}^r
\end{align*}
It emits a load instruction ${\bf lw}\,r\,(k^E)r$ containing
(encrypted)  displacement constant $k{=}{-}\Delta e$
that compensates the offset $\Delta e$ in the address in $r$. 
The processor does the calculation
$a^\E=r \mathop{[-]} \E[\Delta e]$ that produces the encrypted
address $a^\E$ and passes it as-is for lookup
by the memory unit.\footnote{In our own prototype processor for
encrypted computing, a frontend to the address translation
lookaside buffer (TLB) memoises \cite{ishihara06} the encrypted address to
a physically backed sub-range of the full memory address space.
%
The memoisation is
changed randomly at every write through it, so a physical observer
sees a random pattern approximating 
oblivious RAM (ORAM) \cite{ostrovsky1990}.}
%
The entry retrieved from memory has the shared offset
$\Delta{\rm A}$ and the compiler emits a short-form add instruction
${\bf add}\,r\,r\,k^\E$ with semantics
$r{\leftarrow}r{\mathop{[+]}}k^\E$ and $k{=}\Delta' r{-}\Delta {\rm A}$
to change it to a new, freely chosen offset $\Delta' r$ in $r$. The
complete code emitted is:
\begin{align}
\S{D: *e}^r &= (\hbox{\em mc},\Delta)
        \label{e:pointer}\\
        \hbox{\em mc} &= \hbox{\em mc}_e;~
                      \mbox{{\bf lw} r $(\E[-\Delta e])$r}
                      \notag\\
                      &\quad \mbox{{\bf add} r r $\E[\Delta'r-\Delta A]$}
                      \notag
\end{align}
An indexed array lookup A[i] is handled by dereferencing a pointer *(A+i).
Does that follow the principle \eqref{e:maxH}? The add instruction
is varied as the compiler chooses, but the load instruction is not.
However, a load instruction is not an {\em arithmetic} instruction and
\eqref{e:maxH} refers only to those.  A load instruction is a copy from
RAM and should just copy.  Where in RAM the read is physically
mapped to is up to the hardware and should be varied by it
independently. A test of whether two encrypted addresses are equal
based on if they retrieve the same values from RAM does not break
encryption because the lookup is of the encrypted not the decrypted
address.  The general compilation technique for dealing with this
situation (`hardware aliasing'; the term originated in \cite{Barr98})
in which the program has different names for one RAM location
is described in \cite{BB14a,BB14c} (the memory address must be saved for
reuse in reads between consecutive writes, not recalculated; in
particular, the classical frame pointer register is used to save the
stack pointer register on entry to a subroutine and for restoration at
subroutine exit).


Writing an array entry is more problematic, because it should change the
offset delta beneath the encryption.  Because that is shared across the
whole array, therefore every array entry must be rewritten to the new
offset whenever one is written, an {\bf O}($n$) `write storm.' But the
$n-1$ writes to the other array entries all install the same offset.
That contradicts the principle \eqref{e:maxH} that each such arithmetic
write must exercise the possibilities for variation to the maximum. Each
instruction could vary independently, but is constrained by the
convention that the offset $\Delta A$ holds array-wide. Therefore this
`efficient' approach is wrong.  Nevertheless, because it is a
straightforward extension of the integers-only compilation technique, it
is the one presently implemented in our compiler.  Although solo array
reads are more efficient, blinding which array element is really being
read from requires a `read storm' like the write storm, so it is not
more efficient if a compiler codes for that.

\subsection{One Offset per Array  Entry}

\noindent
An array may also be viewed as a single (encrypted) $n\times 32$ bit long
integer variable A, with a single $n\times 32$ bit offset
$\Delta {\rm A}{=}(\Delta{\rm A}_i)_0^{n-1}$ beneath the encryption.
Extending Defn.~\ref{d:2}:
\begin{definition}
Encryption of $n\times32$-bit integers $x$ concatenates the encryptions of 
the 32-bit components:
\begin{align*}
x^\E = \E[x]= \E[x_0\mathop. \dots \mathop. x_{n-1}]&=
   \E[x_0]\mathop. \dots\mathop. \E[x_{n-1}]
\end{align*}
\end{definition}
\noindent
%
The compiler must generate a `write storm' to the whole of the array
after writing one entry and changing its offset delta because it does
not know at compile time which entry A$_i$ (and its associated offset delta
$\Delta$A$_i$)
will be rewritten at runtime, so it must plan to rewrite all -- or
rewrite none, which would go against \eqref{e:maxH}.  Each write in the
write storm contributes new trace information -- the new delta offset -- and
hence entropy.

As stated above, this is the correct approach but our own prototype
compiler does not yet implement it.  The software engineering
perspective is not clear as to whether moving forward to single but long
integer deltas, or multiple 32-bit deltas like those already used for
doubles, is the least difficult development route.  The `single shared
32-but offset' $\Delta$A not $(\Delta$A$_i)$ approach for an array A is
what is currently in use.


\section{{Structs}}
\label{s:Struct}

\noindent
C `structs' are records with fixed fields.  The approach the compiler
takes is to maintain a different offset per field, per variable of
struct type.  That is, for a variable x of struct type with fields\ .a
and\ .b the compiler maintains offsets $\Delta$x.a and $\Delta$x.b.  It
is as though there were two variables, x.a and x.b.

In the case of an array A the entries of which are structs with fields\
.a and\ .b, the compiler maintains two separate sets of offsets
$\Delta$A$_i$.a and $\Delta$A$_i$.b and so on recursively if the fields
are themselves structs.  Updating one field in one entry changes the
associated offset and is accompanied by a `write storm' of adjustments
over the stripe through the array consisting of that same field in all
entries.  That is more efficient than a storm to all fields, so for more
efficient computing in this context, array entries should be split into
structs whenever possible.

\section{{Unions}}
\label{s:Union}

\noindent
The obfuscation scheme in a union type such as
\[\tt
union~\{ struct \{int~a; float~b[2];\};
        double~c[2];
\}
\]
engages compatible offset schemes for the component types.
The offset scheme for the struct will have the pattern (in 32-bit
words) $x, y_0, y_1$, with $x$ the offset for the int and $y_0$, $y_1$
the offsets
for the float array entries, while the pattern for the double
array will be $u_0, v_0, u_1, v_1$.
\begin{align*}
\tt union  \{&\tt struct
\{\underbracket{~\raisebox{0pt}[2ex][0.6ex]{$\tt
int~a$}}_{\raisebox{-1pt}[2ex][0pt]{$x$}}; 
                       \underbracket{\tt float~b[2]}_{\mbox{$y_0,y_1$}};\};
             \underbracket{\tt double~c[2]}_{\raisebox{-1pt}{$u_0,v_0,u_1,v_1$}}; \}
\end{align*}
The resolution is $x{=}u_0{=}\alpha$, $y_0{=}v_0{=}\beta$,
$y_1{=}u_1{=}\gamma$, $v_1{=}\delta$
for a scheme $\alpha,\beta,\gamma,\delta$. That is
the least restrictive obfuscation scheme forced by the union
layout here, and it means that a write to one target field within the
union can be just that.

With our compiler's present (inadequate) solution for arrays,
$y_0{=}y_1$ so $\beta{=}\gamma$, and $u_0{=}u_1$ so $\alpha{=}\gamma$,
and $v_0{=}v_1$ so $\beta{=}\delta$. That gives
$\alpha{=}\beta{=}\gamma{=}\delta$ and the scheme
$\alpha,\alpha,\alpha,\alpha$ of offsets.
That needs a write storm to update the deltas across
the whole union after an update to just one field.
Not only is that inefficient, but it carries no extra
entropy into the trace, contradicting \eqref{e:maxH}.

\section{Theory}
\label{s:Theory}

\noindent
By a {\em trace} $T$ of a program at runtime is meant the sequence of
writes to registers and memory locations.  If a location is read for the
first time without it having previously been written in the trace, then
that is not part of the trace but an {\em input} to it.

Trace $T$ is a random variable, varying from
recompilation to recompilation of the same source code by the compiler.
The compiler freely chooses delta offset schemes for each point in the
code as described in previous sections, and the probability distribution
for $T$ depends on the distribution of those choices.  After a simple
assignment to a register $r$, the trace is longer by one: $T' = T
{}^{\frown} \langle r{=}\E[v] \rangle$.
Let $\H({T})$ be the {\em entropy} of {\em trace} $T$ in this stochastic
setting.  Let $f_T$ be the probability distribution of $T$, then the
entropy is the expectation
\begin{equation}
\H(T)=\mathds{E}[-\log_2 f_{T}]
\end{equation}
The increase in
entropy from $T$ to $T'$ (it cannot decrease as $T$ lengthens) is the
number of bits of unpredictable information added.  A flat distribution
$f_{T}=k$ (constant) uniquely has maximal entropy $\H({T})=\log_2(1/k)$.
Only this fragment of information theory will be required: {\em adding a
maximal entropy signal to a random variable with any distribution at all
on a $n$-bit space gives another maximal entropy, i.e., flat, distribution}.
 
If the offset $\Delta r$ beneath the encryption is chosen randomly and
independently with flat distribution by the compiler, so it has maximal
entropy, then $\H(T')=\H(T)+32$, because there are 32 bits of
unpredictable information added in via the 32-bit delta to the 32-bit
value beneath the encryption, so the 32-bit sum value plus delta
varies with (32-bit) maximal entropy.

Although per instruction the compiler has free choice in accord with
\eqref{e:maxH}, not all the register/memory write instructions issued by
the compiler are jointly free as to the offset delta for the target
location -- it is constrained to be equal at the beginning and end of
a loop, and in general at any point where two control paths join:

\begin{definition}
An instruction emitted by the compiler that adjusts the offset in
location $l$ to a final value common with that in a joining control path
is a {\em trailer} instruction.
\end{definition}
\noindent
Trailer instructions come in {\em sets} for each location $l$ for a
control path join, with one member per path.  Each in the set for $l$ is
last to write to $l$ in a control path before the join.  An example
occurs at {\bf return} from a subroutine.  The final offsets per
location must be the same at all exit points from the subroutine and the
arithmetic instructions that write that make them so make up the trailer
instruction sets.

Because running through the same instruction twice, or a instruction with
the same delta offset for the target location a second time, does not
add any new entropy (the delta offset is already determined for the
second encounter by the first en\-counter), the total entropy in a trace can
be counted as follows:

\begin{lemma}
The entropy of a trace compiled according to \eqref{e:maxH} is $32(n+m)$
bits, where $n$ is the number of distinct arithmetic instructions that
write in the trace, counted once only per set if they are one of a set
of trailer instructions, and once each if they are not, and $m$ is the
number of input words.
\end{lemma}
\noindent
Recall that
`input' is provided by those instructions that read for a first time
in the trace a location not written in it earlier.

Observing data at any point in the trace that has been written by a program
instruction (or read from a location in memory that has not yet been
written) sees variation across recompilations.  The compiler principle
\eqref{e:maxH} guarantees that every opportunity provided by the
emission of an arithmetic instruction that writes is taken by the
compiler as a point at which new variation is introduced.  But at
`trailer' instructions as defined above the compiler jointly organises
several instructions to provide the same final delta to a location and
that is sometimes unnecessary, because that location is never read
again.  Then the variation the compiler has introduced is not maximal,
because it could be increased by varying deltas independently among the
trailer instructions.

To make the trailer instruction synchronisation necessary we consider
that the code might be embedded in any surrounding code, including 
that which reads all locations affected. Then the trailer
synchronisation is necessary and the compiler has done the best job
possible in terms of introducing as much entropy as possible.

\begin{proposition}
The entropy of a program trace compiled according to \eqref{e:maxH}
with synchronisation only at trailer instructions before different
control paths join is maximal over the space of all possible
variations of the constant parameters in the machine code,
given that it works correctly in any context.
\label{t:3}
\end{proposition}
\noindent
The proposition implies a full 32 bits of entropy in the variation 
beneath the encryption must exist in any location at any point in the
trace where the location has been written, or not yet being written,
is read.  The datum in that location has no other option for coming to
be.  This is the  result \eqref{e:S} obtained by structural induction
in \cite{BB17a}:

%
\begin{corollary}
The probability across different compilations by a compiler that
follows principle \eqref{e:maxH}
that any particular 32-bit value has encryption $\E[x]$ in
a given register or memory location at any given point in the program
at runtime is uniformly $1{/}2^{32}$.
\label{t:4}
\end{corollary}

\noindent
That is what formally implies \eqref{e:ddagger}, relative to the
security of the encryption. But a stronger result can now be obtained
from the understanding in the lemma and proposition above:

\begin{definition}
Two data observations in the trace are (delta) {\em dependent} if
they are of the same register at the same point, are input and output
of a copy instruction, or are of the same register 
after the last write to it in a control path before a join and
before the next write.
\label{d:8}
\end{definition}
\noindent
If the trace is observed at two (in general, $n$) independent
points, the variation is maximal possible:
\begin{theorem}
The probability across different compilations by a compiler that
follows principle \eqref{e:maxH} that any $n$ particular
32-bit values in the trace have encryptions $\E[x_i]$, provided they are
pairwise (delta) independent, is $1{/}2^{32n}$.
\label{t:1}
\end{theorem}

\noindent
Each dependent pair reduces the entropy by 32 bits.

\section{Discussion}
\label{s:Discuss}

\noindent
Theorem~\ref{t:1} quantifies exactly the cross-correlation that exists
beneath the encryption in a trace from compiled code where the compiler
is built according to the principle \eqref{e:maxH} (every arithmetic
instruction that writes is varied to the maximal extent possible across
recompilations).  It `names and shames' the points in the trace where
the induced variation is necessarily weak because of the nature of
computation, and statistical influences from the original source code
may show through.  For example, if the code runs a loop summing the same
value again and again into an accumulator, then looking at the
accumulator shows an observer $\E[a+i*b+\delta]$ for a constant offset
$\delta$.  That is an arithmetic series with unknown starting point and
constant step and it is likely to be one of the relatively few
short-stepping paths, and that can be leveraged into a dictionary attack
on the encryption.

A compiler built following the principle \eqref{e:maxH} does as well as
any may to avoid introducing more such weaknesses.  The only way to
eliminate them is to have no loops or branches in the object code.  That
would be a finite-length calculation or unrolled bounded loop with
branches embedded as $t*x+(1-t)*y$ calculations, where $x$ and $y$ are
the potential outcomes from two branches and $t$ is the outcome of a 1/0
test.

With respect to data structures, \eqref{e:maxH} means that each entry of
an array must have its own individually chosen delta offset from
nominal beneath the encryption, and each write to an array must change
them all, as one must change on write and the compiler does not know
which it will be.  The compiler must emit a `write storm'.  Reads too
are necessarily more inefficient than naively may be expected.  Structs
(records with named fields) have different offsets per field, along the
same lines, but the compiler does know which will be accessed, so there
are no write storms.  Unions do force equalities among the delta
offsets of their fields, but they are to be expected from the aliasing
(if it is worthwhile preserving `trick' code -- type punned or aliased,
writing and reading different types -- is another question, but it would
break legacy codes not to).

This document has not touched on short data structures such as short
integers, but it is a problem as their natural variation is small so
they are intrinsically a good subject for dictionary attacks.  With an
abundance of caution, we treat them as integers with random
high bits, and a poor consequence is that strings are loosely
packed.  The text has also not touched unsigned integers, but the
compiler's treatment is the same as for floats -- that is, they are
regarded as being coded as signed integers (with the same bits). The
platform provides primitive arithmetic operations on them in that
coding (encrypted).

The treatment of short integers raises the question of whether extra entropy
could be introduced by changing to 64-bit or 128-bit plaintext words
beneath the encryption, instead of 32-bit, and correspondingly sized
delta offsets from nominal. We believe that is the correct logical
inference. The 32-bit range of variation of standard-sized integers
would be swamped by a 64-bit delta introduced by the compiler and the
looped stepping example $\E[a + i*b + \delta$ above would have a 64-bit
$\delta$, so would have $2^{64}$ possible origin points for the path for
any hypothetical step $b$, not just $2^{32}$, which is too many to
examine in a practical dictionary attack. A 256-bit
encryption for 128-bit plaintext words with 128-bit deltas introduced by
the compiler could be sufficient for all practical purposes  since no
measurement on the trace could then have less than 128 bits of entropy
(Corollary~\ref{t:4} makes this observation).

A particular concern is whether interactions with memory reveal too
much.  One can imagine, for example, testing if two data values are
equal beneath the encryption by seeing if, used as addresses in a load
instruction, they pull the same values into registers.  But load
and store do not resolve the address beneath the encryption. Instead
they pass the literal, encrypted address as-is to the memory unit (which
is not privy to the encryption), so identity of the encrypted addresses
is what would be tested and that is visible already to an observer.
The `hardware aliasing' that multiple encryptions of the same address
causes in use in load and store from the program's point of view is
dealt with by the compiler -- it emits code to save the address verbatim
at first write for subsequent reuse.

At the current stage of development, our own prototype compiler
(\url{http://sf.net/p/obfusc}) has near
total coverage of {\sc ansi} C with GNU extensions, including
statements-as-expressions and expressions-as-statements.  It lacks {\bf
longjmp}, computed goto and global data shared across different
compilation units (a linking issue).

%
%
%

\section{{Conclusion}}

\noindent
How to compile compound and nested C data structures for encrypted
computing extending existing compiler-based `obfuscation' in this
context has been set out here.  A single compiler principle is proposed
-- if any arithmetic instruction that writes is emitted, then it must be
varied by the compiler to the maximal extent possible from recompilation
to recompilation.  Then the compiler is `best possible' in terms of
introducing entropy beneath the encryption in a program runtime trace,
and that is what provides protection against decryption attempts in this
context.  The quantitative theory improves the existing `cryptographic
semantic security relative to the security of the encryption' result for
encrypted computing.


\renewcommand{\baselinestretch}{0.9}
\bibliographystyle{IEEEtran}
\begin{scriptsize}
\bibliography{IEEEabrv,dsn-2019c-nonanon}
\end{scriptsize}
\renewcommand{\baselinestretch}{1}

\end{document}